\documentstyle[aps, manuscript]{revtex}
\tightenlines
\draft
\begin{document}
\title{Decaying cosmological
parameter in the early universe from NKK theory of gravity}
\author{$^{1}$Mauricio Bellini\footnote{
E-mail address: mbellini@mdp.edu.ar}}
\address{$^1$Departamento de F\'{\i}sica, Facultad de
Ciencias Exactas y Naturales,
Universidad Nacional de Mar del Plata and
Consejo Nacional de Ciencia y Tecnolog\'{\i}a (CONICET),
Funes 3350, (7600) Mar del Plata, Argentina.}

\vskip .2cm
\maketitle
\begin{abstract}
Using a formalism recently introduced we study the decaying of the
cosmological parameter during the early evolution of an universe,
whose evolution is governed by a vacuum equation of state.
We use a stochastic approach in a nonperturbative treatment of the
inflaton field from a Noncompact Kaluza-Klein (NKK) theory, to study
the evolution of energy density fluctuations in the early universe.
\end{abstract}
\vskip .2cm                             
\noindent
Pacs numbers: 04.20.Jb, 11.10.kk, 98.80.Cq \\
\vskip .1cm

\section{Introduction}

Cosmological observations imply that there exists an extremely small
upper limit on the vacuum energy density in the present state of our
universe.
This stands in sharp contradiction with theoretical predictions\cite{wei}.
In fact, any mass scale in particle physics contributes to the vacuum energy
density much larger than this upper bound\cite{salehi}. According to modern
quantum field theory, the structure of the vacuum is turned out to be
interrelated with some spontaneous symmetry-breaking effects through
the condensation of quantum scalar fields. This phenomenon gives rise to
a non-vanishing vacuum energy density $\rho_{vac} \sim M^4_p$ ($M_p=G^{-1/2}$
is the Planckian mass and $G$ is the gravitational constant).
The appearance of this characteristic mass scale
may have an important effect on the cosmological constant because it receives
potential contributions from this mass scale due to mass spectrum of
corresponding physical fields in quantum field theory. By taking into account
this contribution, an effective cosmological constant is defined as the
sum of the bare cosmological constant $\lambda$ and
$8\pi G \rho_{vac}$\cite{chen}. This type of contribution gives rise to an
immediate difficulty called the cosmological constant problem. There are
some possible solutions to this problem rendering $\Lambda$ exactly
or almost vanishing\cite{jafari}. One of them consists to find some
relaxation mechanism by which $\Lambda$ could relax to its present
day small value\cite{viana,cald}.
A credible mechanism for obtaining such a decay
already exists, which is to assume the existence of a scalar
field presently relaxing towards the minimum of its potential.
Scalar fields are not only predicted to exist by some particle phsics theories
that go beyond the Standard Model, but are also the most plausible engine
behind a possible inflationary period in the very early
universe\cite{infl1,infl2,infl3,infl4}.
In this work we shall study a possible mechanism for a decaying cosmological
parameter from the KK formalism, but by considering the
extra (spatial-like) dimension $\psi$ as noncompact\cite{c,c1}.
This theory, also called induced-matter theory is, in its simplest
form, the basic KK theory in which the fifth
dimension is not compactified and the field equations of general relativity
in 4D follow from the fact that the 5D manifold is Ricci-flat.
Thus the large
extra dimension is responsible for the appearance of sources in 4D
general relativity. Hence, the 4D world of general relativity is embedded
in a 5D Ricci-flat manifold.
There has recently been an uprising interest in finding exact solutions
of the KK field equations in 5D, where the
fifth coordinate is considered as noncompact. This theory reproduces
and extends known solutions of the Einstein field equations in 4D. Particular
interest revolves around solutions which are not only
Ricci flat, but also Riemann flat. This is because it is
possible to have a flat 5D manifold which contains a curved
4D submanifold, as implied by the Campbell theorem. So, the
universe may be ``empty'' and simple in 5D, but contain matter of
complicated forms in 4D\cite{we}. 

In this work we use a stochastic approach to study the dynamics of the
inflaton field in the early universe,
which is governed by a 4D vacuum
equation of state ${\rm p}_{vac}
= -\rho_{vac} = -{\Lambda\over 8\pi G}$, being
$\Lambda$ the time dependent (decaying) cosmological parameter. To make it
we shall use a 5D canonical metric which is Riemann flat ($R^A_{BCD}=0$)
and describes a 5D apparent vacuum ($G_{AB}=0$). To describe the system
we shall propose a 5D action for a purely kinetic inflaton field which
is minimally coupled to gravity.

\section{5D Formalism}

We consider the
5D canonical metric
\begin{equation}\label{6}
dS^2 = \psi^2 \frac{\Lambda(t)}{3} dt^2 - \psi^2 e^{2 \int
\sqrt{{\Lambda\over 3}}dt} dr^2 - d\psi^2,
\end{equation}
where $dr^2=dx^2+dy^2+dz^2$, being $x,y, z$ dimensionles spatial coordinates. 
Furthermore, $t$ and $\psi$ has spatial units (in this paper
we shall consider $c=\hbar =1$.
We shall assume in what follows that the extra dimension is
spacelike and that the universe is 3D spatially flat, isotropic and
homogeneous.
The metric (\ref{6}) is flat $R^A_{BCD}=0$ and describes a
5D manifold in apparent vacuum ($G_{AB}=0$)
and is a special case 
of the much-studied class of canonical metrics
$dS^2 = \psi^2 g_{\mu\nu} dx^{\mu} dx^{\nu} - d\psi^2$\cite{PDL,MLW,otros}.

To describe neutral matter in a 5D geometrical vacuum
(\ref{6}) we can consider the Lagrangian
\begin{equation}\label{1}
^{(5)}{\rm L}(\varphi,\varphi_{,A}) =
-\sqrt{\left|\frac{^{(5)}
g}{^{(5)}g_0}\right|} \  ^{(5)}{\cal L}(\varphi,\varphi_{,A}),
\end{equation}
where $|^{(5)}g|=\psi^8 (a/a_0)^6$,
is the absolute value of the determinant for the 5D covariant
metric tensor with
components $g_{AB}$ ($A,B$ take the values $0,1,2,3,4$) and
$|^{(5)}g_0|=\psi^8_0$
is a constant of dimensionalization determined
by $|^{(5)}g|$ evaluated at $\psi=\psi_0$ and $a_0=a(t=t_0)$.
Here, $a(t)$ the scale factor
of the universe such that $\dot a/a =\sqrt{\Lambda/3}$.
To describe the system we consider an action
\begin{displaymath}
I = - {\Large\int} d^4x d\psi \sqrt{\left|\frac{^{(5)}g}{^{(5)}g_0}\right|}
\left[\frac{^{(5)} R}{16\pi G} + {\cal L}(\varphi,\varphi_{,A})\right],
\end{displaymath}
where $\varphi$ is a scalar field minimally coupled to gravity.
Furthermore, $^{(5)} R=0$ is the 5D Ricci scalar.

To describe the apparent vacuum, we shall
consider the density Lagrangian
${\cal L}$ in (\ref{1}) must to be
\begin{equation}\label{1'}
^{(5)}{\cal L}(\varphi,\varphi_{,A}) = 
\frac{1}{2} g^{AB} \varphi_{,A} \varphi_{,B},
\end{equation}
for a free scalar field.

The dynamics for $\varphi(t,\vec r, \psi)$ being given by the
equation
\begin{equation}\label{lag}
\ddot\varphi
+ \left( 3\sqrt{\frac{\Lambda}{3}} - \frac{\dot\Lambda}{\Lambda}
\right) \dot\varphi -\frac{\Lambda}{3} e^{-2\int \sqrt{\frac{\Lambda}{3}}dt}
\nabla^2_r\varphi - \frac{\Lambda}{3} \left[
4\psi \frac{\partial\varphi}{\partial\psi} + \psi^2 \frac{\partial^2\varphi}{
\partial\psi^2} \right] =0.
\end{equation}
On the other hand, $\varphi$ complies with the commutation expression
\begin{equation}\label{con}
\left[\varphi(t,\vec r,\psi), \Pi^t(t,\vec{r'},\psi')\right] =
\frac{i}{a^3_0} \  g^{tt} \left|\frac{^{(5)} g_0}{^{(5)} g}\right| \left(
\frac{\Lambda_0}{\Lambda}\right)
\delta^{(3)}(\vec r - \vec{r'}) \  \delta(\psi - \psi'),
\end{equation}
where $\Pi^t = {\partial {\cal L}\over \partial\varphi_{,t}} = {3\over
\Lambda \psi^2} \dot\varphi$ and $a_0$ is the scale factor
of the universe when inflation starts.
As can be demonstrated $\varphi(t,\vec r,\psi)$
$=
e^{-{1\over 2} \int \left[3\left({\Lambda\over 3}\right)^{1/2}
- {\dot\Lambda\over \Lambda}\right]dt} \left(\psi_0/\psi\right)^2
\chi(t,\vec r)$,
so that ${\partial \varphi \over \partial\psi}$ $=-{2\over\psi} \varphi$ and
$\chi$ can be written as a Fourier expansion
\begin{equation}\label{fou}
\chi(t,\vec r) = \frac{1}{(2\pi)^{3/2}} {\Large\int} d^3k_r
{\Large\int} dk_{\psi} \left[ a_{k_r k_{\psi}}
e^{i(\vec{k_r}.\vec{r} +\vec{k_{\psi}}.\vec{\psi})}
\xi_{k_r k_{\psi}}(t,\psi)
+
a^{\dagger}_{k_r k_{\psi}}
e^{-i(\vec{k_r}.\vec{r} +\vec{k_{\psi}}.\vec{\psi})}
\xi^*_{k_r k_{\psi}}(t,\psi)\right],
\end{equation}
such that
\begin{equation}\label{comu}
\left[ \chi(t,\vec{r}),\dot\chi(t,\vec{r'})\right] =
\frac{i}{a^3_0} \  \delta^{(3)}(\vec{r}
-\vec{r'}) \  \delta(\vec{\psi} - \vec{\psi'}),
\end{equation}
and $\xi_{k_r k_{\psi}}(t,\psi) = e^{-i \vec{k_{\psi}}.\vec{\psi}}
\bar\xi_{k_r k_{\psi}}(t)$. The commutator (\ref{comu}) is satisfied
for $\left[a_{\vec{k_r}k_{\psi}}, a^{\dagger}_{\vec{k_r'}k_{\psi}'}\right]=
\delta^{(3)}(\vec{k_r} - \vec{k_r'}) \  \delta(\vec{k_{\psi}}-\vec{k_{\psi}'})$
and  $\left[a^{\dagger}_{\vec{k_r}k_{\psi}}, a^{\dagger}_{\vec{k_r'}k_{\psi}'}
\right]=
\left[a_{\vec{k_r}k_{\psi}}, a_{\vec{k_r'}k_{\psi}'}\right]=0$, if the
following condition holds:
\begin{equation}\label{cond}
\bar\xi_{k_r k_{\psi}}(t) \dot{\bar\xi}^*_{k_r k_{\psi}}(t) -
\bar\xi^*_{k_r k_{\psi}}(t) \dot{\bar\xi}_{k_r k_{\psi}}(t) = \frac{i}{a^3_0}.
\end{equation}

\section{4D dynamics}

To describe the 4D dynamics we
can make a foliation on $\psi=\psi_0$ on the line element
(\ref{6}), such that the effective 4D metric holds:
$\left.dS^2\right|_{eff} = ds^2$, where
\begin{equation}\label{61}
ds^2 = 
\psi^2_0 \frac{\Lambda(t)}{3} dt^2 - \psi^2_0 e^{2 \int
\sqrt{{\Lambda\over 3}}dt} dr^2.
\end{equation}
In this section we shall study the dynamics described by the
inflaton field $\varphi$, making emphasis on the long wavelength section,
which describes this field on cosmological scales.

\subsection{4D effective dynamics of $\varphi$}

The 5D Lagrangian density (\ref{1'}) can be expanded as
\begin{equation}
^{(4)}{\cal L} = \left.\frac{1}{2} g^{\mu\nu} \varphi_{,\mu}\varphi_{,\nu} +
\frac{1}{2} g^{\psi\psi} \varphi_{,\psi}\varphi_{,\psi}\right|_{\psi_0},
\end{equation}
such that the 4D potential is
\begin{equation}
V(\varphi) = \left.
-\frac{1}{2} g^{\psi\psi} \varphi_{,\psi}\varphi_{,\psi}\right|_{\psi_0}=
\frac{2}{\psi^2_0}\varphi^2(t,\vec r,\psi_0).
\end{equation}
Furthermore, from the equation (\ref{lag}), we obtain the effective
4D dynamics for $\varphi(t,\vec r, \psi_0)$
\begin{equation}\label{lag1}
\ddot\varphi + 
\left( 3\sqrt{\frac{\Lambda}{3}} - \frac{\dot\Lambda}{\Lambda}
\right) \dot\varphi -\frac{\Lambda}{3} e^{-2\int \sqrt{\frac{\Lambda}{3}}dt}
\nabla^2_r\varphi - \left.\frac{\Lambda}{3} \left[
4\psi \frac{\partial\varphi}{\partial\psi} + \psi^2 \frac{\partial^2\varphi}{
\partial\psi^2} \right]\right|_{\psi_0} =0,
\end{equation}
where we can make the identification
\begin{equation}
V'(\varphi) =
- \left.\frac{\Lambda}{3} \left[
4\psi \frac{\partial\varphi}{\partial\psi} + \psi^2 \frac{\partial\varphi}{
\partial\psi^2} \right]\right|_{\psi_0} = \frac{2\Lambda}{3}
\varphi(t,\vec r,\psi_0).
\end{equation}
After make the transformation $\varphi(t,\vec r,\psi=\psi_0) =
e^{-{1\over 2} \int \left[3\left({\Lambda\over 3}\right)^{1/2}
- {\dot\Lambda\over \Lambda}\right]dt}
\chi(t,\vec r)$, we obtain
the equation of motion for the redefined field
$\chi$
\begin{equation}\label{eqm}
\ddot\chi - \frac{\Lambda}{3} e^{-2
\int \left(\frac{\Lambda}{3}\right)^{1/2} dt}
\nabla^2_r\chi -\left[\frac{\Lambda}{12} + \frac{3\dot\Lambda^2}{4\Lambda^2} -
\frac{\ddot\Lambda}{2\Lambda}\right] \chi =0,
\end{equation}
which can be expanded on the hypersurface $\psi=\psi_0$, as
\begin{equation}
\chi(t,\vec r) = \frac{1}{(2\pi)^{3/2}} {\Large\int}d^3k_r {\Large\int}
dk_{\psi} \left[ a_{k_rk_{\psi}} e^{i \vec{k_r}.\vec{r}}
\bar\xi_{k_rk_{\psi}}(t) + 
a^{\dagger}_{k_rk_{\psi}} e^{-i \vec{k_r}.\vec{r}}
\bar\xi^*_{k_rk_{\psi}}(t)\right] \  \delta(k_{\psi} - k_{\psi_0}).
\end{equation}
Furthermore, the modes $\bar\xi_{k_rk_{\psi}}(t)$ are given
by the equation of motion
\begin{equation}\label{motion}
\ddot{\bar\xi}_{k_r k_{\psi_0}} + \left[ \frac{k^2_r \Lambda}{3}
e^{-2\int \left(\frac{\Lambda}{3}\right)^{1/2} dt} -
\left(\frac{\Lambda}{12} +\frac{3\dot\Lambda^2}{4 \Lambda^2}
- \frac{\ddot\Lambda}{
2\Lambda}\right)\right] \bar\xi_{k_r k_{\psi_0}} =0.
\end{equation}

\subsection{4D stochastic dynamics of $\chi$
on cosmological scales}

In order to describe separately the long and short wavelength sectors
of the field $\chi$
we can define the fields $\chi_L(t,\vec r)$
and $\chi_S(t,\vec r)$
\begin{eqnarray}
\chi_L(t, \vec r) &=& \frac{1}{(2\pi)^{3/2}} {\Large\int} d^3 k_r
{\Large\int} \  \theta(\epsilon k_0(t) - k_r) \left[
a_{k_r k_{\psi}} e^{i \vec{k_r}.\vec{r}} \bar\xi_{k_r k_{\psi_0}}(t)
+ c.c.\right]
\delta(k_{\psi} - k_{\psi_0}), \\
\chi_S(t, \vec r) &=& \frac{1}{(2\pi)^{3/2}} {\Large\int} d^3k_r
{\Large\int} \  \theta(k_r- \epsilon k_0(t)) \left[
a_{k_r k_{\psi}} e^{i \vec{k_r}.\vec{r}} \bar\xi_{k_r k_{\psi_0}}(t) + c.c.\right]
\delta(k_{\psi} - k_{\psi_0}),
\end{eqnarray}
where $c.c$ denotes the complex conjutate and
$k_0(t) = e^{\int \sqrt{\Lambda/3} dt}\left[
{1\over 4} + {9\over 4} {\dot\Lambda^2\over \Lambda^3} -
{3\ddot\Lambda\over 2\Lambda^2}\right]^{1/2}$. The field
that describes the dynamics of $\chi$ on the infrared sector
($k^2_r \ll k^2_0$) is $\chi_L$. Its dynamics obeys
the Kramers-like stochastic equation
\begin{equation}
\ddot\chi_L - \frac{k^2_0}{a^2} \chi_L =
\epsilon \left[ \frac{d}{dt}\left(\dot{k}_0 \eta(t,\vec r) \right)+
\dot{k}_0 \gamma(t,\vec r)\right],
\end{equation}
where ${k^2_0\over a^2} = \left({3\over \Lambda_0}\right) \left[
{1\over 4} + {9\over 4} {\dot\Lambda^2\over \Lambda^3} - {3\ddot\Lambda\over
2\Lambda^2}\right]^{1/2}$
and the stochastic operators $\eta$, $\kappa$ and $\gamma$ are
\begin{eqnarray}
\eta &=& \frac{1}{(2\pi)^{3/2}} {\Large\int} d^3 k_r \  \delta(\epsilon k_0-k_r)
\left[a_{k_r k_{\psi_0}} e^{i \vec{k_r}.\vec{r}} \bar\xi_{k_r k_{\psi_0}}(t)
+ c.c.\right],\\
\gamma &=& \frac{1}{(2\pi)^{3/2}}
{\Large\int} d^3 k_r \  \delta(\epsilon k_0-k_r)
\left[a_{k_r k_{\psi_0}} e^{i \vec{k_r}.\vec{r}} \dot{\bar\xi}_{k_r
k_{\psi_0}}(t)
+ c.c.\right].
\end{eqnarray}
This second order stochastic equation can be rewritten as two
coupled stochastic equations
\begin{eqnarray}
&& \dot u = \frac{k^2_0}{a^2} \chi_L
+ \epsilon \dot k_0 \gamma,\\
&& \dot\chi_L = u + \epsilon \dot k_0 \eta,
\end{eqnarray}
where we have introduced the auxiliar field
$u=\dot\chi_L - \epsilon \dot{k}_0 \gamma$.
The condition to can neglect the noise $\gamma$ with respect to
$\eta$, is
\begin{equation}\label{ju}
\frac{\dot{\bar\xi}_{k_R} \dot{\bar\xi}^*_{k_R}}{
\bar\xi_{k_R} \bar\xi^*_{k_R}} \ll \frac{\left(\ddot{k}_0\right)^2}{
\left( \dot{k}_0\right)^2}.
\end{equation}
The Fokker-Planck equation for $P(\chi^{(0)}_L,u^{(0)}|\chi_L,u)$ is
\begin{equation}
\frac{\partial P}{\partial t} = - u \frac{\partial P}{\partial \chi_L}
-\frac{k^2_0}{a^2} \chi_L \frac{\partial P}{\partial
u} + D_{11}(t) \frac{\partial^2 P}{\partial\chi^2_L},
\end{equation}
where $D_{11}(t) = {\epsilon^3\dot k_0 k^2_0\over
4\pi^2 }\left|\bar\xi_{\epsilon k_0}\right|^2$
and $P(\chi^{(0)}_L,u^{(0)}|\chi_L,u)$ describes the probability of
transition from a configuration $(\chi^{(0)}_L,u^{(0)})$ to
$(\chi_L,u)$. Furthermore, $\epsilon \simeq 10^{-3}$ is a dimensionless
constant such that on cosmological scales holds $k_r/k_0 < \epsilon$.
Hence, the equation of motion for $\left< \chi^2_L\right> =
\int d\chi_L du \chi^2_L P(\chi_L,u)$ will be
\begin{equation}\label{mot1}
\frac{d}{dt}\left<\chi^2_L\right> = D_{11}(t) .
\end{equation}
To describe the dynamics of the squared $\varphi_L$-expectation value
we return to the original field $\varphi_L =
e^{-{1\over 2}\int \left[3\sqrt{\Lambda/3} - {\dot\Lambda\over \Lambda}
\right]dt} \chi_L$, so that the equation (\ref{mot1}) can be
rewritten as
\begin{equation}
\frac{d}{dt}\left<\varphi^2_L\right> = -\left[3\sqrt{\frac{\Lambda}{3}}
-\frac{\dot\Lambda}{\Lambda}\right] \left<\varphi^2_L
\right> + D_{11}(t)
e^{-\int \left[3\sqrt{\frac{\Lambda}{3}}
-\frac{\dot\Lambda}{\Lambda}\right] dt},
\end{equation}
which has the following solution
\begin{equation}
\left<\varphi^2_L\right> =
e^{-\int \left[3\sqrt{\frac{\Lambda}{3}}
-\frac{\dot\Lambda}{\Lambda}\right] dt} \left[
\left<\varphi^2_L\right>_0 + {\Large\int} D_{11}(t) dt\right],
\end{equation}
where $\left<\varphi^2_L\right>_0$ is a constant of integration.

In order to understand better this result in the context
of the inflaton field fluctuations $\phi(\vec r, t)$, we
can make the following semiclassical approach:
\begin{equation}
\varphi(\vec r, t) = \left< \varphi(\vec r,t)\right> + \phi(\vec r,t),
\end{equation}
where $\left<\varphi\right> = \phi_c(t)$ and
$\left<\phi\right>=0$.
With this representation one obtains
\begin{equation}
\left<\varphi^2\right> = \phi^2_c + \left<\phi^2\right>,
\end{equation}
where $\phi_c(t)$ is the solution of the zero mode equation in (\ref{lag1})
\begin{equation}\label{phi_c}
\ddot\phi_c + \left[3\sqrt{\frac{\Lambda}{3}} -\frac{\dot\Lambda}{\Lambda}
\right]  \dot\phi_c + \frac{2\Lambda}{3} \phi_c =0.
\end{equation}
The solutions of physical interest for $\phi_c(t)$
should be decreasing with time, so that
after inflation ends
$\phi_c(t\rightarrow \infty) \rightarrow 0$. Hence,
after inflation one obtains the following result:
\begin{equation}
\left<\varphi^2\right>_{{t\over t_0}\gg 1} \simeq
\left<\phi^2\right>_{{t\over t_0}\gg 1},
\end{equation}
which means that for ${t\over t_0} \gg 1$ the following
approximation is fulfilled:
\begin{equation}
\left<\phi^2_L\right>_{{t\over t_0} \gg 1} \simeq
\left<\varphi^2_L\right>_{{t\over t_0} \gg 1} \simeq 
e^{-\int \left[3\sqrt{\frac{\Lambda}{3}}
-\frac{\dot\Lambda}{\Lambda}\right] dt} \left[
\left<\varphi^2_L\right>_0 + {\Large\int} D_{11}(t) dt\right].
\end{equation}

Furthermore,
we can estimate the amplitude of density energy fluctuations on cosmological
scales
\begin{equation}
\left.\frac{\delta\rho}{\rho}\right|_{IR, (t/t_0) \gg 1} \simeq
\frac{\left<V'(\varphi)\right>}{\left<V(\varphi)\right>} \  \left<
\phi^2_L\right>^{1/2}. 
\end{equation}

\section{An example for a time dependent $\Lambda$}

As an example we consider a
cosmological parameter $\Lambda(t) = 3 p^2 t^{-2}$, which is related
to a Hubble parameter $H(t) = \left({\Lambda(t)\over 3}\right)^{1/2} =p/t$.
The effective 4D equation of state is ${\rm p}_{vac} = -\rho_{vac} =
-{\Lambda \over 8\pi G}$. 
In this case the equation of motion (\ref{motion})
for the time dependent modes
$\bar\xi_{k_r k_{\psi_0}}(t)$ is
\begin{equation}
\ddot{\bar\xi}_{k_r k_{\psi_0}} + \frac{1}{t^2} \left[
k^2_r p^2 \left(\frac{t}{t_0}\right)^{-2p} - \frac{p^2}{4}
\right] \bar\xi_{k_r k_{\psi_0}}=0.
\end{equation}
The general solution for this equation is
\begin{equation}\label{gensol}
\bar\xi_{k_r k_{\psi_0}}(t) = 2\nu \sqrt{t/t_0} \left[
\alpha \Gamma(\nu) \left(\frac{t}{t_0}\right)^{-p\nu} \left(\frac{x(t)}{2}
\right)^{-\nu} {\cal J}_{\nu}[x(t)] + \beta \Gamma(-\nu) \left(\frac{t}{t_0}
\right)^{p\nu} \left(\frac{x(t)}{2}\right)^{\nu} {\cal J}_{-\nu}[x(t)]
\right],
\end{equation}
where $\alpha$ and $\beta$
are constants of integration, $\nu = {\sqrt{p^2+1}\over 2p}$
and $x(t) = k_r \left(t_0/t\right)^p$.
In this general case the solution is very difficult to be normalized by
the expression (\ref{cond}) on the hypersurface $\psi=\psi_0$. However,
for $p=2n$ ($n$ integer positive), one obtains
\begin{equation}
\bar\xi_{k_rk_{\psi_0}}(t) = \sqrt{\frac{t}{t_0}} \left[
A_1 \  {\cal H}^{(1)}_{\nu}[x(t)] + B_1 \  {\cal H}^{(2)}_{\nu}[x(t)]\right],
\end{equation}
which can be normalized if we use the Bunch-Davies vacuum\cite{bd}. For
$A_1=0$ and $a_0={t_0\over p}$,
we obtain $B_1=i {p\over t_0} \sqrt{{\pi\over 8}}$,
so that for $p=2n$ the time dependent modes
$\bar\xi_{k_r k_{\psi_0}}(t)$ holds
\begin{equation}\label{modes}
\bar\xi_{k_r k_{\psi_0}}(t) = i \frac{p}{2 t_0}
\sqrt{\frac{\pi t}{2 t_0}} \  {\cal H}^{(2)}_{\nu}[x(t)].
\end{equation}

On long wavelength modes ($k_r \ll (t/t_0)^p$), the
second kind Hankel function ${\cal H}^{(2)}_{\nu}[x(t)]$ takes the
asymptotic form $\left.{\cal H}^{(2)}_{\nu}[x(t)]\right|_{x\ll 1} \simeq
{-i\over \pi} \Gamma(\nu) \left({x\over 2}\right)^{-\nu}$, so that
the modes are
\begin{equation}\label{solut}
\left.\bar\xi_{k_r k_{\psi_0}}(t)\right|_{k_r \ll (t/t_0)^p}
\simeq  \frac{p}{2 t_0} \sqrt{\frac{t}{2\pi t_0}} \Gamma[\nu]
\left[\frac{k_r}{2} \left(\frac{t_0}{t}\right)^p\right]^{-\nu}.
\end{equation}
The diffusion coefficient $D_{11}(t)$ in this particular case
is
\begin{equation}
D_{11}(t) = \frac{\epsilon^{3-2\nu} p^3 \Gamma^2[\nu] 2^{4(\nu-2)}}{
\pi^3 t^3_0} \left(\frac{t}{t_0}\right)^{3p},
\end{equation}
so that, for late times (i.e., at the end of inflation) one obtains
\begin{equation}
\left<\phi^2_L\right>_{IR,(t/t_0) \gg 1}
\simeq \left(\frac{t}{t_0}\right)^{-(3p+2)}
\left[\left<\varphi^2_L\right>_0 + \frac{
\epsilon^{3-2\nu} p^3 \Gamma^2[\nu] 2^{4(\nu-2)}}{
\pi^3(1+3p) t^2_0}
\left(\frac{t}{t_0}\right)^{3p+1} \right],
\end{equation}
which is valid on cosmological scales. For late times one obtains
$\left<\phi^2_L\right>_{IR} \sim t^{-1}$, independently of the power $p$.
The evolution of $\phi_c(t)$ for this model is
\begin{equation}
\phi_c(t) = \phi_0 \left(\frac{t}{t_0}\right)^{-\left(\frac{1+3p}{2}\right)}
\left[ A \  \left(\frac{t}{t_0}\right)^{\frac{\sqrt{1+6p+p^2}}{2}} +
B \  \left(\frac{t}{t_0}\right)^{-\frac{\sqrt{1+6p+p^2}}{2}}\right],
\end{equation}
where $\phi_0$ is $\phi_c(t=t_0)$ and $(A,B)$ are dimensionless constants
such that $A+B=1$.
Note that $\phi_c$ is monotonically decreasing.
Finally, the energy density fluctuations for late times are
\begin{equation}
\left.\frac{\delta\rho}{\rho}\right|_{IR,(t/t_0) \gg 1 } \simeq
\frac{\phi_c(t)}{\left<\phi^2_L\right>} \left<\phi^2_L\right>^{1/2},
\end{equation}
which for very large $p$ go as $
\left.{\delta\rho \over \rho}
\right|_{IR,(t/t_0) \gg 1, p\gg 1 } \sim t^{-p}$.\\

\section{Finnal Comments}

We have studied a cosmological model for the early universe
from a NKK
theory of gravity with a space-like extra dimension,
where the
cosmological parameter decreases with time.
We have worked a stochastic treatment for the
effective 4D inflaton field
without the hypothesis of a slow - roll regime.
Hence, the dynamics the field on large scales is described
by a second order stochastic equation.
In this framework the long - wavelength modes
of the inflaton field reduces to a quantum system subject to a quantum
noise which is originated by the short - wavelength sector.
In this approach, the effective 4D potential is quadratic in $\varphi $
and has a geometrical origin. As in STM theory\cite{we} of gravity
4D source terms are induced from a 5D vacuum and the
fifth dimension (here a space-like one) is noncompact.
In our theory the 5D
vacuum is represented by a 5D globally flat metric (which
describes a 5D apparent vacuum $G_{AB}=0$) and a purely
kinetic density Langrangian for a quantum scalar field minimally coupled
to gravity.

In the example here studied for $\Lambda = 3p^2 t^{-2}$, we obtain
that the energy density fluctuations decrease monotonically with time.
In particular, we obtain that for very large $p$ these fluctuations
go as $t^{-p}$.\\

\vskip .3cm
\noindent
M.B. Acknowledges CONICET and UNMdP (Argentina) for financial
support.\\

\end{document}